\RequirePackage{fix-cm}
\documentclass[smallcondensed]{svjour3}     
\smartqed  
\usepackage{graphicx}
\usepackage{url}
\usepackage{amsmath}
\usepackage{txfonts}
\usepackage{natbib}
\newcommand{\teff}{$T_{\rm eff}$}
\newcommand{\tts}{$T_{\rm S}$}
\newcommand{\ttp}{$T_{\rm P}$}
\newcommand{\fff}{$ff$}
\newcommand{\ddt}{${\rm\Delta}T$}

\begin{document}
\graphicspath{{images/}}

\title{EChO spectra and stellar activity II. The case of dM stars.
}


\author{Gaetano Scandariato \and 
	Giuseppina Micela}


\institute{G. Scandariato \at
              INAF - Osservatorio Astrofisico di Catania\\
              via S. Sofia, 78, 95123 Catania\\
              \email{gas@oact.inaf.it}           
           \and
           G. Scandariato \and G. Micela \at
              INAF - Osservatorio Astronomico di Palermo \lq\lq G. Vaiana\rq\rq
}

\date{Received: date / Accepted: date}

\maketitle

\begin{abstract}

EChO is a dedicated mission to investigate exoplanetary atmospheres. When extracting the planetary signal, one has to take care of the variability of the hosting star, which introduces spectral distortion that can be mistaken as planetary signal. Magnetic variability is a major deal in particular for M stars. To this purpose, assuming a one spot dominant model for the stellar photosphere, we develop a mixed observational-theoretical tool to extract the spot's parameters from the observed optical spectrum. This method relies on a robust library of spectral M templates, which we derive using the observed spectra of quiet M dwarfs in the SDSS database. Our procedure allows to correct the observed spectra for photospheric activity in most of the analyzed cases, reducing the spectral distortion down to the noise levels. Ongoing refinements of the template library and the algorithm will improve the efficiency of our algorithm.

\keywords{Exoplanets \and Atmosphere \and Stellar Activity \and EChO}
\end{abstract}

\section{Introduction}\label{sec:intro}

The flux variations caused by magnetic activity can hamper the extraction of the exoplanet atmosphere signal and a need arises to diagnose stellar variability mostly in the near-IR and mid-IR continuum. Such variations are associated with active regions (cool and/or bright spots) coming on and off view as the star rotates and from intrinsic variability of such active regions \citep{Ribas2014}. Furthermore, the impact of activity varies with wavelength and, during transmission spectroscopy, it may be confused as a planetary atmosphere effect \citep{Ballerini2012}.

These variations can occur on relatively many timescales compared to the planet’s orbital period, and thus impact directly on EChO's observation strategy, which will consist in the combination of different epochs of eclipse data \citep{Tinetti2012}.

The problem of magnetic activity is a main issue in particular for M dwarfs, which may be largely covered by photospheric cool spots \citep{Berdyugina2005}. In the recent years some work has been done using optical spectra, but focusing on the chromospheric indicators \citep[see, e.g., ][]{Cincunegui2007, Martinez2011, Gomes2011, Stelzer2013}, while there is poor knowledge on the infrared, mostly due to lack of data. Moreover, if the star hosts hot Jupiters, then planetary atmospheres may be as hot as photospheric spots, thus affecting the same spectral ranges.

We can assume that the transiting planetary atmosphere is almost transparent in the optical, its opacity showing molecular features at wavelengths longer than 2.5$\mu$m \citep{Tinetti2012}. One possible approach is thus to obtain simultaneous optical and infrared spectra, using the former as a calibrator to correct the effects of activity. Observations in the visible range are thus essential to provide the stellar data needed for the measurement and interpretation of exoplanet atmospheres. EChO will adopt such approach.

From the theoretical point of view, synthetic models of dM spectra still present some inconsistencies compared to the observations, due to the difficulties in implementing realistic molecular opacities, grain formation and convection in the atmosphere \citep{Allard2011}. Finally, may theoretical models perfectly reproduce the observed spectra, the robustness of the method described in \citet{Micela2014} decreases with later spectral types.

To overcome all these difficulties, in this paper we discuss a new mixed observation-model approach for the analysis and correction of the effects of magnetic activity in dM spectra. In particular, in Sect.~\ref{sec:coaddition} we build an empirical library of quiet M spectra using the database of the Sloan Digital Sky Survey (SDSS) \citep{York2000}, while in Sect.~\ref{sec:model} we discuss the results of our spectral fitting algorithm.

\section{Empirical spectral templates}\label{sec:coaddition}

We take advantage of the catalog of 70,841 M spectra selected by \citet{West2011} from the SDSS. These spectra, which are flux-calibrated by the SDSS pipeline, cover the wavelength range from $\sim$3800 Å to $\sim$9000 Å with a spectral resolution of R$\sim$2000. \citet{West2011} also provide a number of measurements relative to the spectra (e.g.\ radial velocity, interstellar extinction, metallicity index, H$_{\rm \alpha}$ equivalent width\dots), which we use to discard outliers and/or peculiar stars.

We use these measurements to select good representatives of quiet M stars in the solar neighborhood. We will carefully describe this selection in a forthcoming paper dedicated to the study of activity in low mass stars. Here we remark that our first selection includes bona-fine disk stars with solar-like metallicity and poor interstellar extinction.Then, for each subtype we select 25\% of the spectra with the lowest emission in H$_{\rm\alpha}$, which are likely the quietest spectra in the selected subsamples. Finally, for each subsample we keep 66\% of the spectra with the highest S/N. The number of selected spectra for each subtype is reported in Table~\ref{tab:samples}.

\begin{table}
\begin{center}
\caption{Number of selected spectra for each spectral subtype used to build the templates of quiet stars.}\label{tab:samples}
\begin{tabular}{ccr}
 \hline\hline
Subtype & T$_{\rm eff}$ (K)\footnote{Temperature scale from \citet{Reid2005}.} & Number of spectra\\
\hline
M0 & 3800 & 111\\
M1 & 3600 & 97\\
M2 & 3400 & 126\\
M3 & 3250 & 148\\
M4 & 3100 & 121\\
M5 & 2800 & 57\\
M6 & 2600 & 94\\
M7 & 2500 & 102\\
M8 & 2400 & 23\\
M9 & 2300 & 6\\
\hline
\end{tabular}
\end{center}
\end{table}

After correction for Doppler effect, we degrade the spectra down to EChO's nominal spectral resolution (R=300), propagating the SDSS flux uncertainties accordingly. Then, we scale the spectra (both fluxes and corresponding uncertainties) such that the integrated flux over the 5500$\div$9000~\AA\ range matches the flux of the synthetic spectrum computed by \citet{Allard2011} with the same temperature. We adopt the temperature scale of \citet{Reid2005}. 

\begin{figure*}
\centering
\includegraphics[viewport=0 0 1440 720,clip,width=\linewidth]{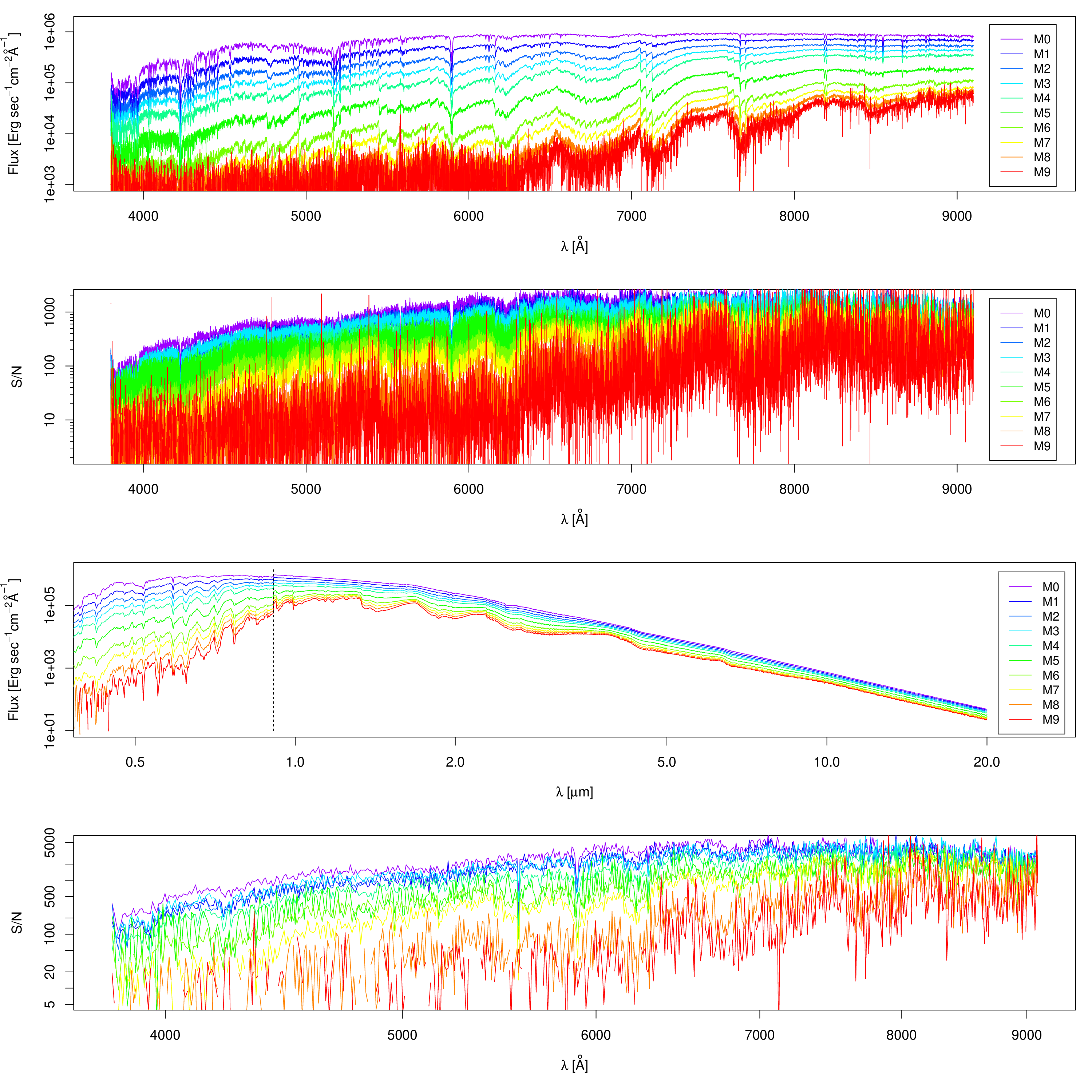}
\caption{\textit{Top panel - }Spectral templates covering the optical and infrared ranges with spectral resolution R=300. The vertical dashed line marks the conjunction between the SDSS spectra and the BT-Settl synthetic model. \textit{Bottom panel - } S/N of the template spectra shown in the top panel.}\label{fig:spectraE}
\end{figure*}

Once the spectra are justified in wavelength and luminosity, for each spectral subtype we compute the sigma-clipped average weighted over the flux uncertainties, obtaining our quiet templates. These are extended over the infrared range (up to $\lambda$=20$\mu$m) using the synthetic spectra of \citet{Allard2011} with the same temperature. Our spectral templates are shown in Fig.~\ref{fig:spectraE}.

\section{Correcting for photospheric activity}\label{sec:model}

In this Section we describe our algorithm to correct the distortion in M dwarfs' spectra introduced by photospheric active regions using the spectral templates derived in Sect.~\ref{sec:coaddition}.

The proposed approach is based on the following assumptions \citep{Micela2014}:
\begin{itemize}
\item photospheric activity is due to the presence of a dominant spot cooler than the photosphere, i.e.\ \tts$<$\ttp, covering a fraction \fff\ of the stellar surface (\ttp\ is the temperature of the unspotted photosphere, \tts\ and \fff\ are the spot's temperature and filling factor respectively);
\item the observed flux $F_{observed}$ can be expressed as:
\begin{equation}
F_{observed}=(1-ff)\cdot F_P+ff\cdot F_S,\label{eq:flux}
\end{equation}
where $F_P$ and the $F_S$ are the fluxes radiated by the unspotted photosphere and the spot respectively.
\item the spectrum radiated by a M star is largely determined by its photospheric temperature, while the presence of active regions on the stellar surface introduces a first order distortion in the spectrum. We thus assume that spectral typing is weakly affected by photospheric activity.
\end{itemize}

We use our spectral templates to model the flux $F_P$ from the unperturbed photosphere, at temperature \ttp, and the flux $F_S$ radiated by the spot, at temperature \tts. We extend our analysis down to the M7 spectral type, due to the fact that the M8 and M9 templates descend from statistically poor samples ($\leq$25 spectra, Table~\ref{tab:samples}) and, by consequence, are quite uncertain, with S/N$\sim$200 over the analyzed spectral range.

We focus on the 5500~\AA$\div$9000~\AA\ spectral range, entirely covered by the SDSS spectra: the blue cutoff is presumably the bluest wavelength observable with EChO, while the red cutoff is a compromise between the need to have a band broad enough to include the spectral features sensitive to activity, and maintain as small as possible the overlap with the band \lq\lq interesting\rq\rq\ for planetary atmosphere observations. In this spectral range we exclude a few narrow windows bracketing the chromospheric lines, i.e.\ the Na\textsc{i} doublet (not resolved at EChO's resolution, at $\sim$5893~\AA), the H$_{\rm\alpha}$ line ($\lambda$6562.79), the Ca\textsc{ii} IRT ($\lambda$8498.02, $\lambda$8542.09, $\lambda$8662.14), the CaOH band head at $\sim$6225~\AA\ \citep{Bochanski2007}.

Since we foresee the application of this method also for stars with no information on the distance, in the following analysis we will normalize the spectra over the integrated spectra in the analyzed spectral range (5500~\AA$\div$9000~\AA\ excluding the small gaps discussed above).

\subsection{The algorithm for the spot's parameters retrieval}\label{sec:algorithm}

We analyze the effect of activity separately for each subtype. We fix the subtype, and we build a 2D grid of spotted spectra in the \fff$\times$\ddt\ space, where \ddt=\ttp-\tts. In this space, \fff\ ranges in the [0,0.5] interval with a step of 0.01, while \ddt\ runs from 50~K up to the temperature difference corresponding to a M7 spot, with a step of 20~K.
  For each node in the grid, we compute the corresponding spotted spectrum using Eq.~\ref{eq:flux}. This grid is thus a bidimensional collection of reference spectra for the following fitting procedure.



Then, we simulate a number of spotted spectra with different combinations of \fff\ and \ddt\ (not corresponding to any of the grid nodes), over which we run our fitting algorithm aimed at recovering the simulated parameters.

\begin{figure}
\centering
\includegraphics[width=\linewidth]{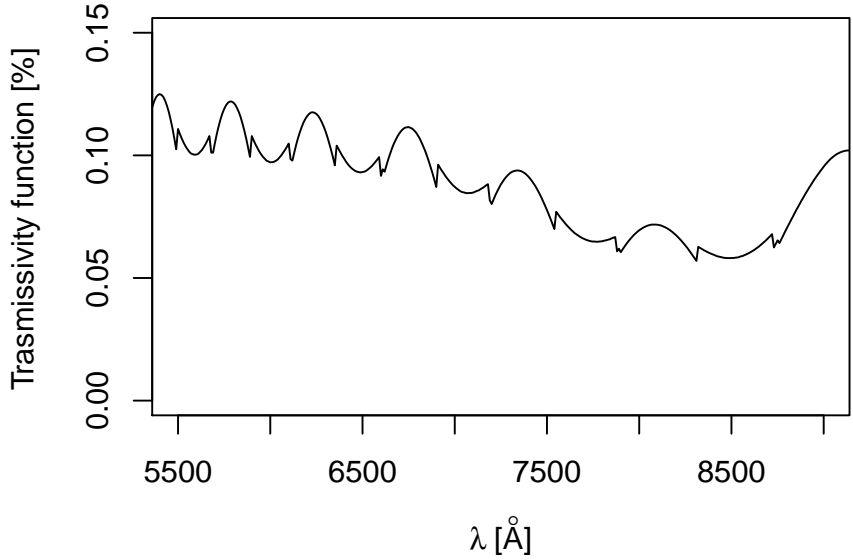}
\caption{EChO's transmissivity function in the analyzed spectral range (Adriani, private communication).}\label{fig:efficiency}
\end{figure}

We also simulate the noise associated with the observed spectra using the transmissivity function delivered by EChO's consortium (Adriani, \textit{private communication}, Fig.~\ref{fig:efficiency}), available up to 2.5~$\rm\mu$m. In particular, the noise we simulate is such to have S/N$_{8200\AA}$=200 per resolution element, which is the minimum requirement for the EChO mission.

In order to recover the simulated parameters, we fit the observed spectrum using the Nelder-Meade (or downhill simplex) method, which minimizes the weighted $\chi^2$ over the \ddt\ and \fff\ ranges covered by the 2D grid. The minimization method is such that the solution is searched over the continuous parameter intervals, i.e.\ it is not constrained on the grid nodes (as for the nearest neighbor method).

\subsection{Results from the algorithm}\label{sec:results}

\begin{figure*}
\centering
\includegraphics[viewport=278 -10 556 277,clip,width=.3\linewidth]{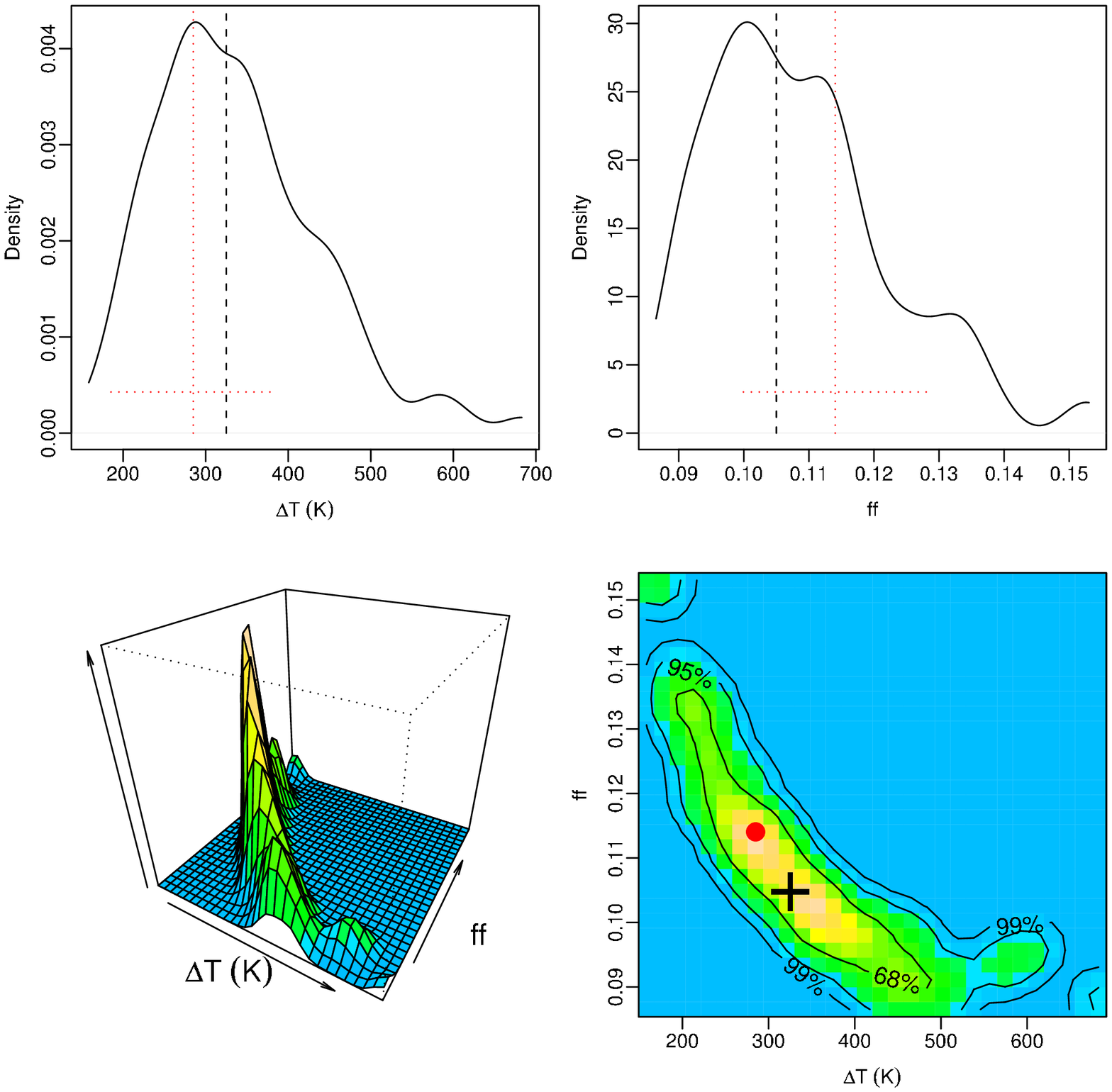}
\hspace{1cm}
\includegraphics[viewport=0 0 553 275,clip,width=.6\linewidth]{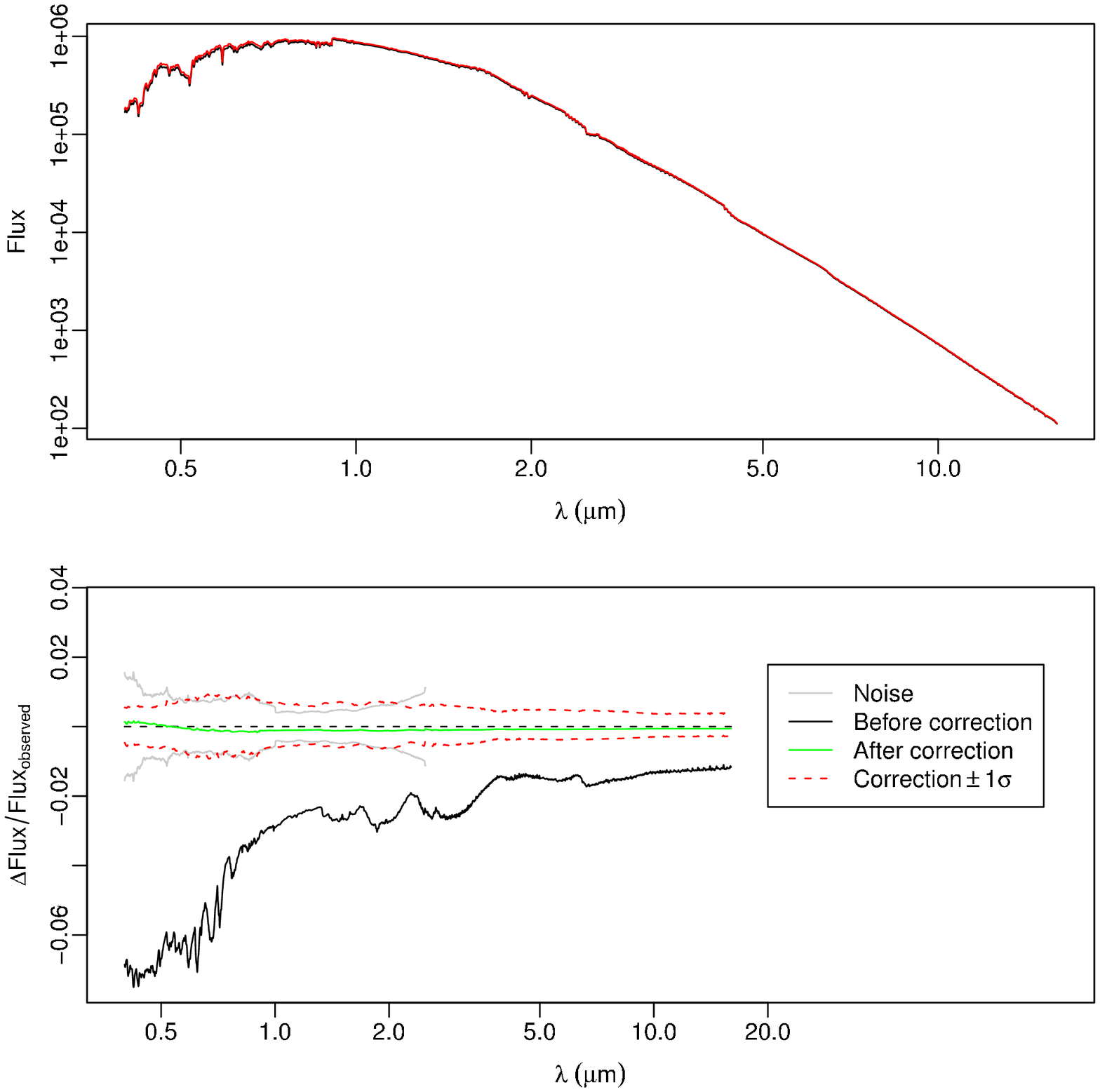}
\caption{Statistical analysis of the results from our fitting algorithm run over 1000 realizations of noisy spectra. The simulated case is a M0 star with a spot \ddt=325~K cooler than the photosphere and with \fff=0.10. \textit{Left panel - }2D probability distribution function of the fitted parameters. The black cross represents the input parameters, while the red dot marks the mode of the distribution. The contours show the confidence regions around the mode containing the 68\%, 95\% and 99\% of the sample. \textit{Right panel - } Correction of the observed spectrum. The black solid line is the systematic error made if the observed spectrum is corrected with the corresponding quiet template. The gray lines show the confidence band corresponding to the simulated S/N (no noise model is available for $\lambda>$2.5$\mu$m). The green line is the residual distortion if the spectrum is corrected with the best-fit spotted template (red spot in the left panel). The dashed red lines delimit the envelop of the solutions contoured by the 68\% level in the left panel.}\label{fig:bestfit}
\end{figure*}

To test the robustness of our algorithm, we generate 1000 random realizations of flux noise to associate to the simulated spectrum, and we run the fitting algorithm on each noisy spectrum. In Fig.~\ref{fig:bestfit} we show the output of our algorithm for the test case of a M0 star with a spot \ddt=325~K cooler than the photosphere and with \fff=0.10. The mode of the 2D probability distribution function of the 1000 best-fits (left panel) is consistent with the simulated parameters within 1 sigma. In particular, we remark that the geometry of the confidence region is such that the the output parameters are anti-correlated, i.e.\ the spot's temperature decreases with increasing coverage. This is consistent with the fact that the impact of the spot in the observed fluxes increases with both parameters. We also remark that the half-width of the \teff\ range spanned by the confidence region is as large as $\simeq$100~K.

In Fig.~\ref{fig:bestfit} we also plot the distortion introduced in the spectrum by the spot, compared with the simulated S/N. The black solid line represents the error one makes when subtracting the quiet template to the observed spectrum. This residual spectrum contains structures above the noise level which significantly affect the planetary signal.

If we correct the observed spectrum using the best-fit parameters, the structures in the residual spectrum (green line) are largely reduced below noise. Moreover, the envelope of the solution within the 68\% confidence region (red dashed lines, see left panel in Fig.~\ref{fig:bestfit}) are consistent with the noise level, indicating that the algorithm introduces negligible uncertainty compared with flux noise.

We tested a number of cases with different spectral types and spot's parameters, obtaining results similar to what is discussed above. To determine the applicability limits of our procedure, we also tested the some challenging test cases.

\paragraph{M0 star, spot with \tts=3475~K and \fff=0.03.}

\begin{figure*}
\centering
\includegraphics[viewport=278 -10 556 277,clip,width=.3\linewidth]{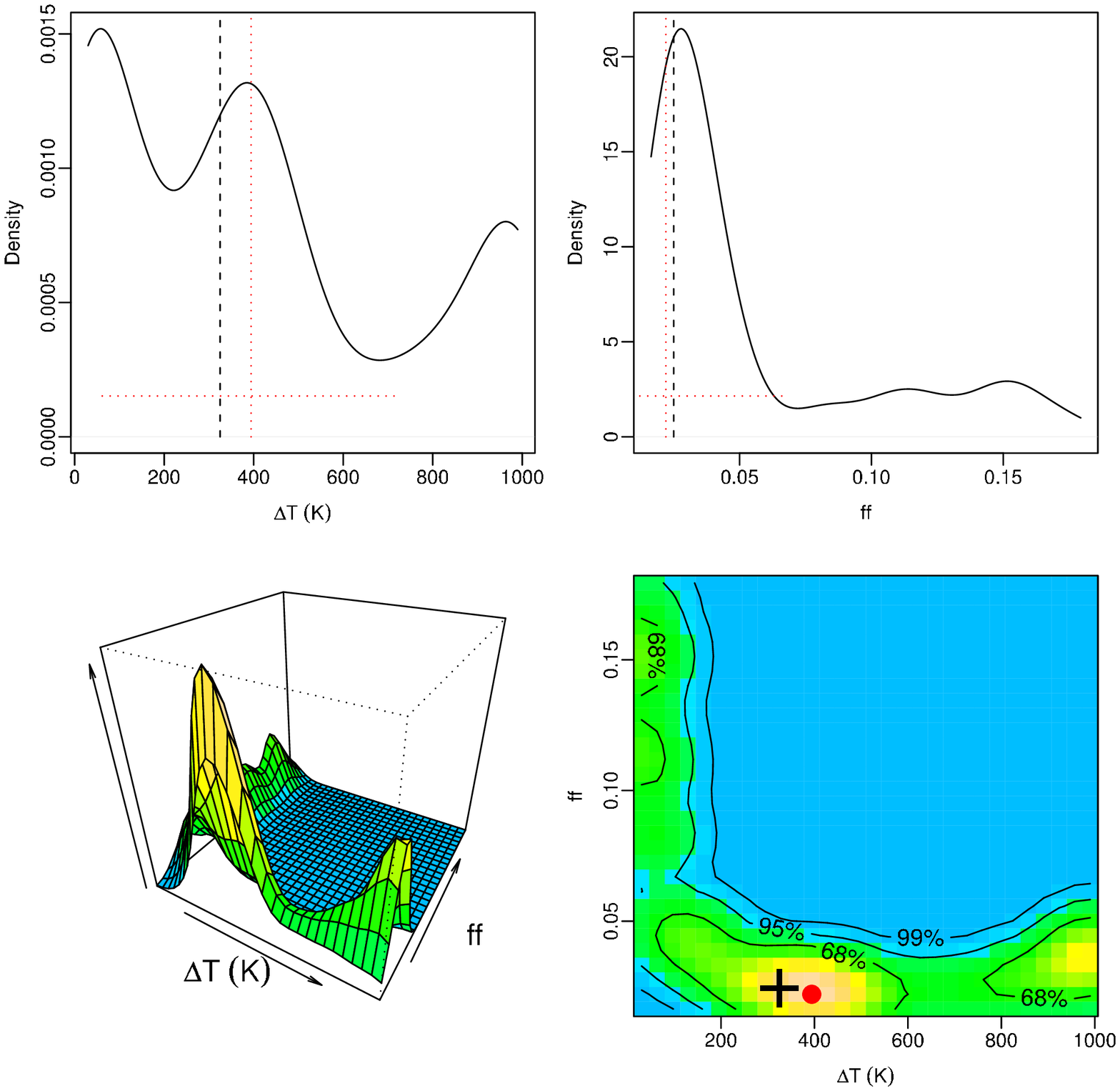}
\hspace{1cm}
\includegraphics[viewport=0 0 553 275,clip,width=.6\linewidth]{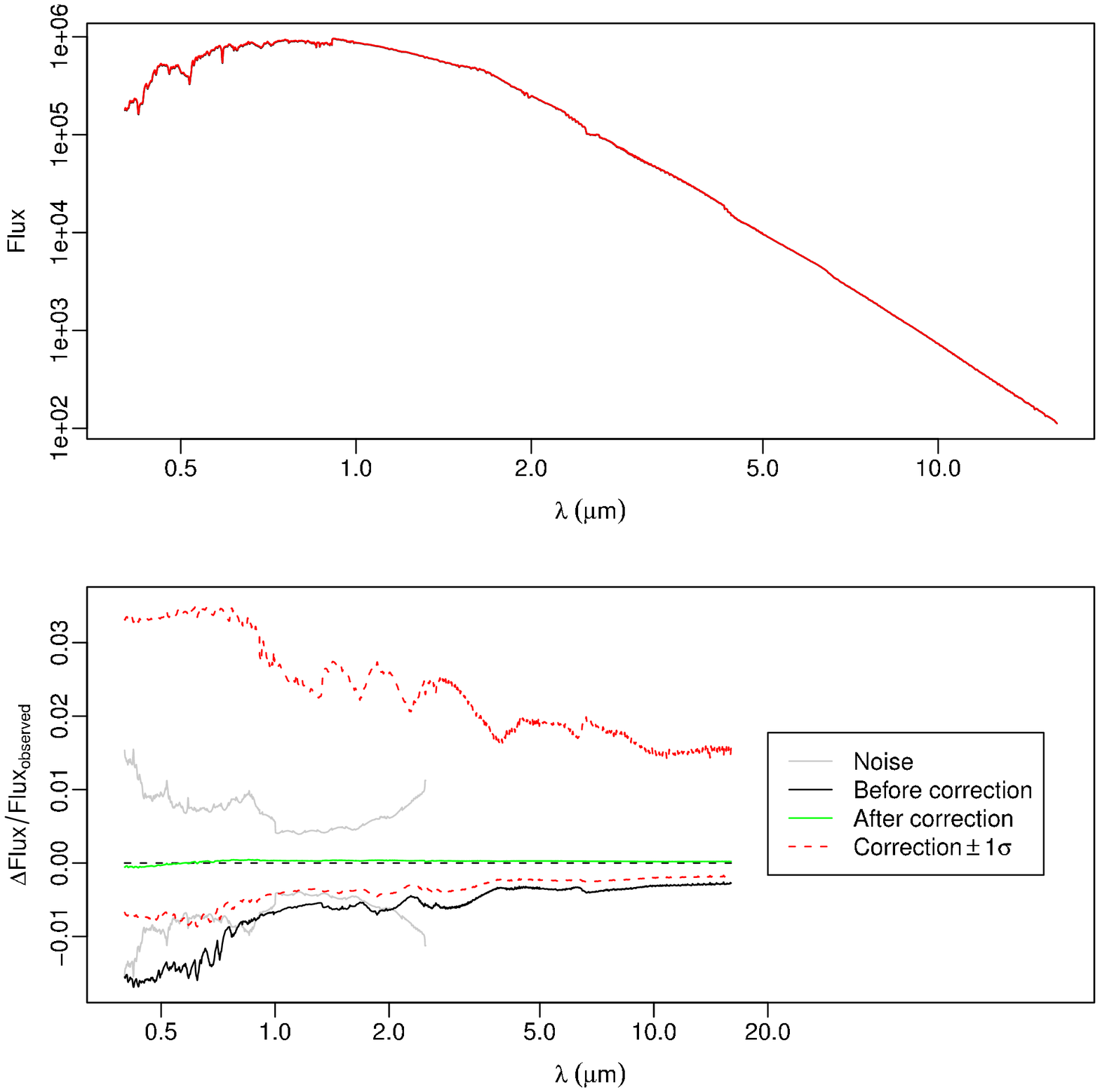}
\caption{Same as in Fig.~\ref{fig:bestfit}, with \fff=0.03.}\label{fig:testcase2}
\end{figure*}

In Fig.~\ref{fig:testcase2} we show the output of the same case discussed above, but with a smaller \fff\ of 0.03. We find that the 1-$\sigma$ 2D region (left panel) spans larger ranges of \fff\ and \tts, including in particular consistent overestimates of the two parameters. This is due to the fact that, since the spectral distortion is comparable with noise, the best fit sometimes collapses to the edge of the investigated parameter space, leading to solutions which overcorrect the observed spectrum above the noise level (right panel).

We find that this is a regular feature for the cases with spectral distortion comparable with noise, due to the fact that the best fit solution tend to collapse at the edge of the investigated parameter space. We are currently working on a way to discriminate and automatically avoid such cases.

Still, the algorithm is able to confidently recover, from a statistical point of view, the simulated parameters such to correct the spectrum within the noise level. The main difference with the previous case is that the wider confidence region pushes the upper 1 $\sigma$ envelope in the right panel well above the simulated noise.

\paragraph{M0 star, spot with \tts=3075~K.}

\begin{figure*}
\centering
\includegraphics[viewport=278 -10 556 277,clip,width=.3\linewidth]{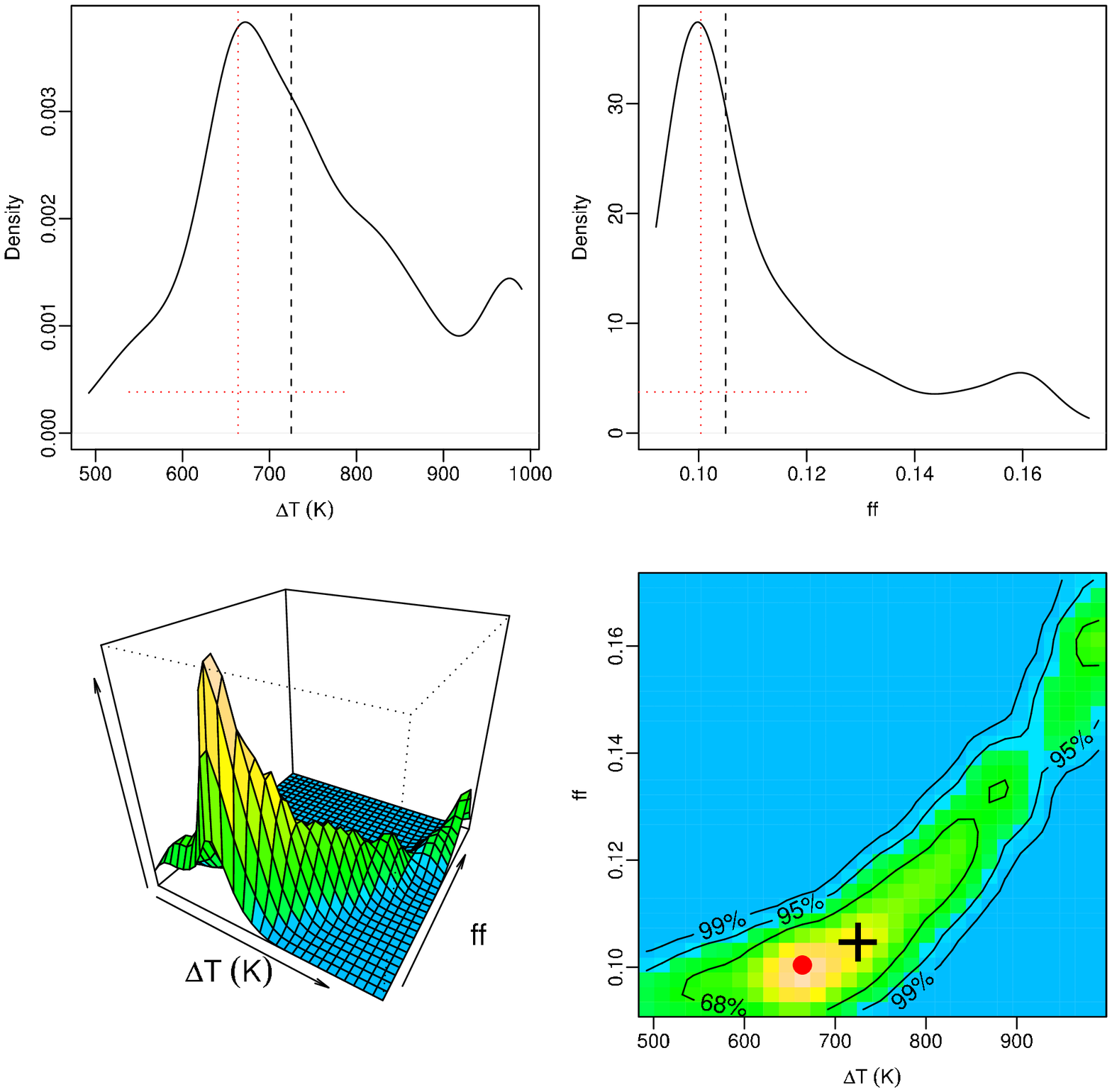}
\hspace{1cm}
\includegraphics[viewport=0 0 553 275,clip,width=.6\linewidth]{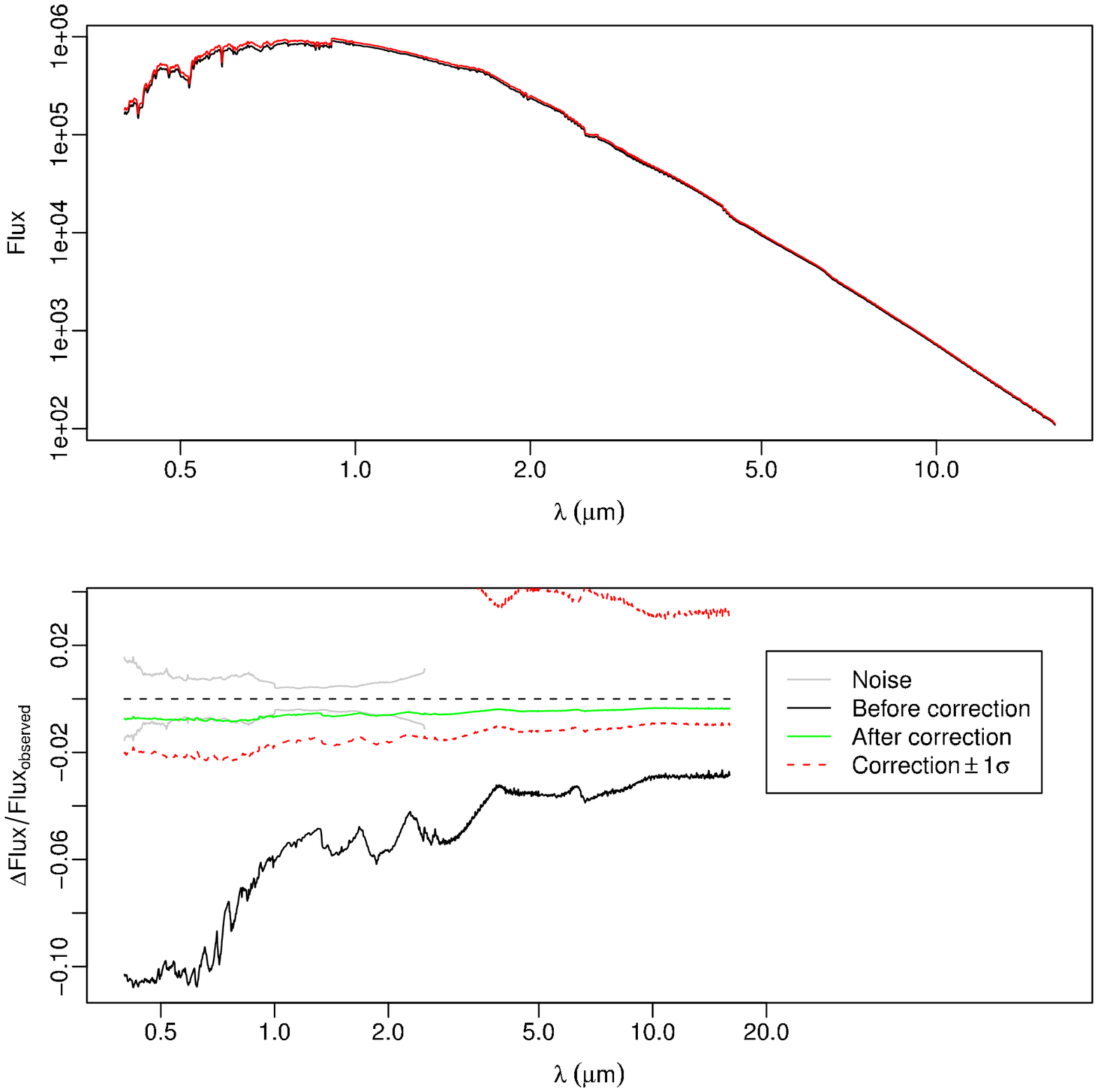}
\caption{Results of the algorithm for a M0 star with \fff=0.10 and \tts=3075~K.}\label{fig:testcase3}
\end{figure*}

In Fig.~\ref{fig:testcase3} we discuss the results for a M0 star with spot temperature \tts=3075~K and \fff=0.10. The effect of such a low temperature is the change of the slope of the confidence region in the 2D plane (left panel), i.e.\ the ensemble of noisy spectra is consistent with spotted M0 photospheres whose temperature and filling factor are directly correlated.

\begin{figure*}
\centering
\includegraphics[width=\linewidth]{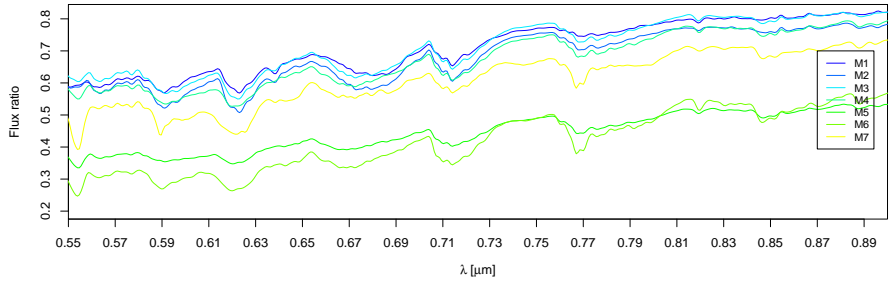}
\caption{Flux ratio between each M subtype and the earlier one. The color code is reported in the legend. The case of M8 and M9 subtypes are not shown to avoid cluttering.}\label{fig:ratio}
\end{figure*}

A careful investigation of this effect shows that this is due to the shape of late M spectra over the analyzed spectral range. As a matter of fact, in Fig.~\ref{fig:ratio} we show the flux ratio between each subtype and the earlier one. We find that at early subtypes (M4 and hotter) the ratio has features related to the functional dependence of molecular bands profiles on \teff. Conversely, at later subtypes the ratios tend to flatten both in general slope and features, i.e.\ late M subtypes roughly differ by just a rescaling factor with poor reshaping of the flux distribution. Thus, the flux-rescaled spectrum of a spot cooler than \teff$\simeq$3100~K may be mimicked by large cool spots or small warm ones. This leads to uncertainties on \tts\ of the order of $\simeq$150~K.

\begin{figure*}
\centering
\includegraphics[viewport=278 -10 556 277,clip,width=.3\linewidth]{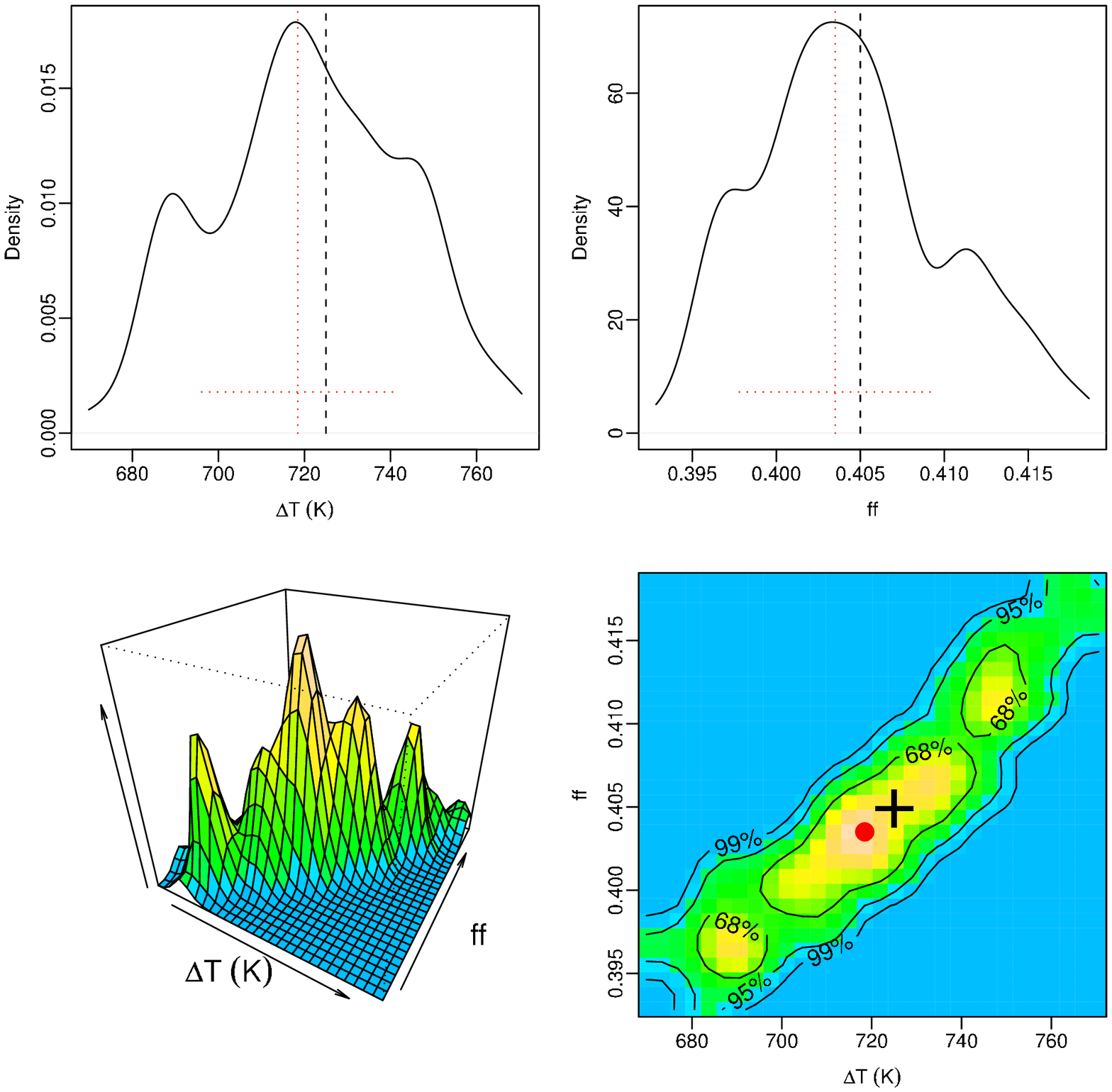}
\hspace{1cm}
\includegraphics[viewport=0 0 553 275,clip,width=.6\linewidth]{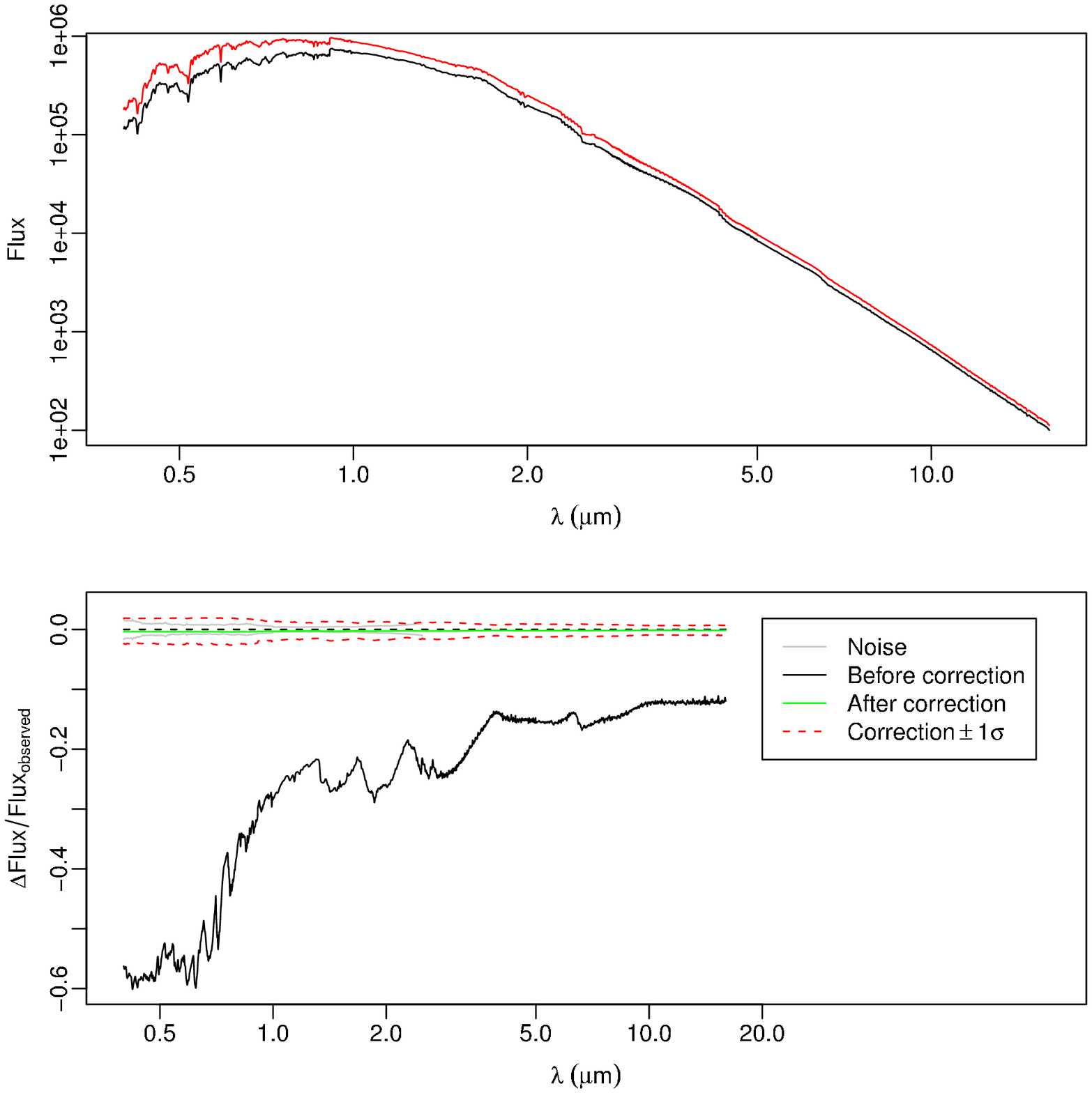}
\caption{Same as in Fig.~\ref{fig:testcase3}, with \fff=0.41.}\label{fig:testcase4}
\end{figure*}

This sort of degeneracy between \tts\ and \fff\ is particularly evident at small filling factors. If the spot coverage increases, the extent of the degeneracy (and the width of the confidence band in the plots) decreases, as shown in Fig.~\ref{fig:testcase4}.

\section{Summary and conclusions}\label{sec:discussion}

In this work we assemble an empirical optical spectral library for M stars coadding the dM spectra in the SDSS database classified by \citet{West2011}, grouped by spectral subtype. We extend the templates to the infrared band assuming the spectral model of \citet{Allard2011}.

For each subtype, we also assemble a set of 1-spot dominant spectra by linearly combining the corresponding template with cooler templates and different filling factors (Eq.~\ref{eq:flux}), thus mimicking the presence of photospheric spots. Then, we develop and test an algorithm to fit the spotted spectra aiming at recovering the spot's temperature and filling factor.

The results show that at the foreseen EChO's S/N we generally recover the spot's parameters such to rectify the spectral distortions introduced by photospheric activity. Moreover, the uncertainty on the corrected optical spectra in most cases is comparable with spectral noise.

Some caution is needed in case of poor contrast between the photosphere and the spot, i.e.\ when the difference between the observed spectra and the template with the same spectral type is comparable to noise. In these cases our algorithm may erroneously diverge to the edges of the investigated parameter space. Despite the correction of such a low distortion, compared with the noise level, may lead to wrong results, we remark that the correction itself is generally of little utility, as the correction is of the same order of magnitude of noise.

Finally, when the spot is cooler than T$\simeq$3100~K, our algorithm is less accurate, in particular at \fff$\lesssim$0.10, as the uncertainty on the correction is generally larger than spectral noise.

Currently, we are working on different aspects of our algorithm trying to achieve better results. We are refining the selection of dM stars to improve our template library. In particular, we are analyzing in details the activity signatures of the stars aiming at (i) a subset of spectra with as negligible as possible activity and (ii) a subset of active spectra to be fitted with our algorithm. This analysis will also indicate the spectral range and activity range over which the algorithm works at best.

The same analysis will also lead to the systematic characterization of photospheric activity in M stars, providing valuable information on the typical temperature and size of spots, eventually depending on spectral subtype. This study will thus provide tools to set boundaries and priors on the fitting algorithm, shrinking the searched parameter space and improving the spot's parameters retrieval. 

We are also working to improve the fitting procedure, in particular the instability at low spot vs.\ photosphere contrast. These will allow us to better constrain the uncertainty on the spectral correction. Nonetheless, we remark that the analysis of the out-of-transit spectra of each EChO's target will likely provide tighter constraints on the spot's parameters, i.e.\ the star itself will suggest the solution to look for and the parameter space to be searched.

\begin{acknowledgements}
The authors acknowledge fundings from the ASI-INAF agreement \lq\lq Missione EChO: assessment phase\rq\rq\ I/022/12/0.

The authors acknowledge A.\ Adriani for providing EChO's transmissivity function.

The authors acknowledge publisher Springer. The final publication of this paper is available at link.springer.com.

Funding for SDSS-III has been provided by the Alfred P. Sloan Foundation, the Participating Institutions, the National Science Foundation, and the U.S. Department of Energy Office of Science. The SDSS-III web site is \url{http://www.sdss3.org/}.

SDSS-III is managed by the Astrophysical Research Consortium for the Participating Institutions of the SDSS-III Collaboration including the University of Arizona, the Brazilian Participation Group, Brookhaven National Laboratory, Carnegie Mellon University, University of Florida, the French Participation Group, the German Participation Group, Harvard University, the Instituto de Astrofisica de Canarias, the Michigan State/Notre Dame/JINA Participation Group, Johns Hopkins University, Lawrence Berkeley National Laboratory, Max Planck Institute for Astrophysics, Max Planck Institute for Extraterrestrial Physics, New Mexico State University, New York University, Ohio State University, Pennsylvania State University, University of Portsmouth, Princeton University, the Spanish Participation Group, University of Tokyo, University of Utah, Vanderbilt University, University of Virginia, University of Washington, and Yale University.
\end{acknowledgements}



\begin{thebibliography}{}

\bibitem[Allard et al.(2011)]{Allard2011} Allard, F., Homeier, D., Freytag, B.:  Model Atmospheres From Very Low Mass Stars to Brown Dwarfs. 16th Cambridge Workshop on Cool Stars, Stellar Systems, and the Sun,  448,  91 (2011) 

\bibitem[Ballerini et al.(2012)]{Ballerini2012} Ballerini, P., Micela, G., Lanza, A.~F., Pagano, I.:  Multiwavelength flux variations induced by stellar magnetic activity: effects on planetary transits. Astronomy and Astrophysics,  539,  A140 (2012) 

\bibitem[Berdyugina (2005)]{Berdyugina2005} Berdyugina, S.~V.:  Starspots: A Key to the Stellar Dynamo. Living Reviews in Solar Physics,  2,  8 (2005) 

\bibitem[Bochanski et al.(2007)]{Bochanski2007} Bochanski, J.~J., West, A.~A., Hawley, S.~L., Covey, K.~R.:  Low-Mass Dwarf Template Spectra from the Sloan Digital Sky Survey. The Astronomical Journal,  133,  531 (2007)

\bibitem[Cincunegui et al.(2007)]{Cincunegui2007} Cincunegui, C., D{\'{\i}}az, R.~F., Mauas, P.~J.~D.:  H{$\alpha$} and the Ca II H and K lines as activity proxies for late-type stars. Astronomy and Astrophysics,  469,  309 (2007) 

\bibitem[Gomes da Silva et al.(2011)]{Gomes2011} Gomes da Silva, J., Santos, N.~C., Bonfils, X., et al.:  Long-term magnetic activity of a sample of M-dwarf stars from the HARPS program. I. Comparison of activity indices. Astronomy and Astrophysics,  534,  A30 (2011) 

\bibitem[Mart{\'i}nez-Arn{\'a}iz et al.(2011)]{Martinez2011} Mart{\'i}nez-Arn{\'a}iz, R., L{\'o}pez-Santiago, J., Crespo-Chac{\'o}n, I., Montes, D.:  Effect of magnetic activity saturation in chromospheric flux-flux relationships. Monthly Notices of the Royal Astronomical Society,  414,  2629 (2011) 

\bibitem[Micela (2014)]{Micela2014} Micela, G.: This Volume of Experimental Astronomy

\bibitem[Reid \& Hawley (2005)]{Reid2005} Reid, I.~N., Hawley, S.~L.. New Light on Dark Stars Red Dwarfs, Low-Mass Stars, Brown Stars.~Praxis Publishing Ltd, (2005) 

\bibitem[Ribas et al.(2014)]{Ribas2014} Ribas, I.: This Volume of Experimental Astronomy

\bibitem[Stelzer et al.(2013)]{Stelzer2013} Stelzer, B., Frasca, A., Alcal{\'a}, J.~M., et al.: X-shooter spectroscopy of young stellar objects. III. Photospheric and chromospheric properties of Class III objects. Astronomy and Astrophysics, 558,  A141 (2013) 

\bibitem[Tinetti et al.(2012)]{Tinetti2012} Tinetti, G., Beaulieu, J.~P., Henning, T., et al.:  EChO. Exoplanet characterisation observatory. Experimental Astronomy,  34, 311 (2012) 

\bibitem[West et al.(2011)]{West2011} West, A.~A., Morgan, D.~P., Bochanski, J.~J., et al.: The Sloan Digital Sky Survey Data Release 7 Spectroscopic M Dwarf Catalog. I. Data. The Astronomical Journal,  141,  97 (2011) 

\bibitem[York et al.(2000)]{York2000} York, D.~G., Adelman, J., Anderson, J.~E., Jr., et al.: The Sloan Digital Sky Survey: Technical Summary. The Astronomical Journal, 120,  1579 (2000) 

\end{thebibliography}
\end{document}